# DOOML: A NEW DATABASE & OBJECT-ORIENTED MODELING LANGUAGE FOR DATABASE-DRIVEN WEB APPLICATION DESIGN AND DEVELOPMENT

Thijs Otter

Faculty of Engineering, Design & Computing,
Inholland University of Applied Sciences, Haarlem, The Netherlands

## ABSTRACT

*A database driven web application is a very common software solution to rising business problems. Modeling the database and the software architecture can be challenging, hence there not being one combined modeling language for database and software architecture, specifically suited for web application development. In this paper we present Database object-oriented Modeling Language (DooML) and its primary Archetype Diagram: a notation for specifying the design of a database schema and corresponding object-oriented software architecture. It combines the syntax for drawing Entity Relationship Diagrams, the Relational Model and Universal Modeling Language Class Diagrams as well to create a mixed diagram, stating database design as well as software design specifications. By default, DooML ensures that the approach of advanced web application development is model-driven and both database-oriented as well as object-oriented.*

## KEYWORDS

*modeling language, object-oriented modeling, database modeling, object-oriented database modeling, archetype diagram, software modelling.*

## 1. INTRODUCTION

In this paper we present a new modeling language for the database-driven development of object oriented (web) applications, that combines both the database-oriented as well as the object-oriented approach of web application development.

The Database Object-Oriented Modeling Language (abbreviated as DooML) is designed to be a universal back-end software engineering modeling language for the purpose of graphically defining a combined model for both database design and object-oriented structural code design. It can be used to design code structures for database-driven web applications that make use of modern object-oriented programming languages (e.g. Java, C# or PHP) and a relational database for data storage.

In modern software development, object-oriented software applications can be modeled using technical diagrams that are part of the UML standard, using class diagrams, sequence diagrams and state diagrams.

For database design, several diagrams can be used, of which the Entity-Relationship Diagram is a commonly used approach. Song, Evans & Park [1] stated several notational techniques that exist to model Entity-Relationship Diagrams, where for the purpose of clarity the method of Chen (1976) is used as inspiration for the development of the DooML modeling language.

DooML combines the syntax for drawing Entity Relationship Diagrams [2], the Relational Model [3] and Universal Modeling Language Class Diagrams [4] or some versions of SysML [5] as well to create a mixed diagram, stating database design as well as software design specifications. It differs from SecureUML [6], as DooML is not emphasizing on creating applications with a focus on security-driven development. The DooML language as such differs from the combined concept of Chen and Lu [7], which is primarily focused on manufacturing control systems (embedded appliances).

The authors have also taken interest into the modeling language called WebML, as described by Ceri, Fraternali and Bongio [8], that is primarily focused on the navigation within a web application but does not make use of a components that describe the (relational) data storage.

Figure 1 shows common modeling principles for each of the development areas: application modeling (e.g. UML), database modeling (using ERD or EER) or API modeling (using SwaggerHub). A combined database/application/API modeling language should reside in the center of the Venn-diagram: DooML.

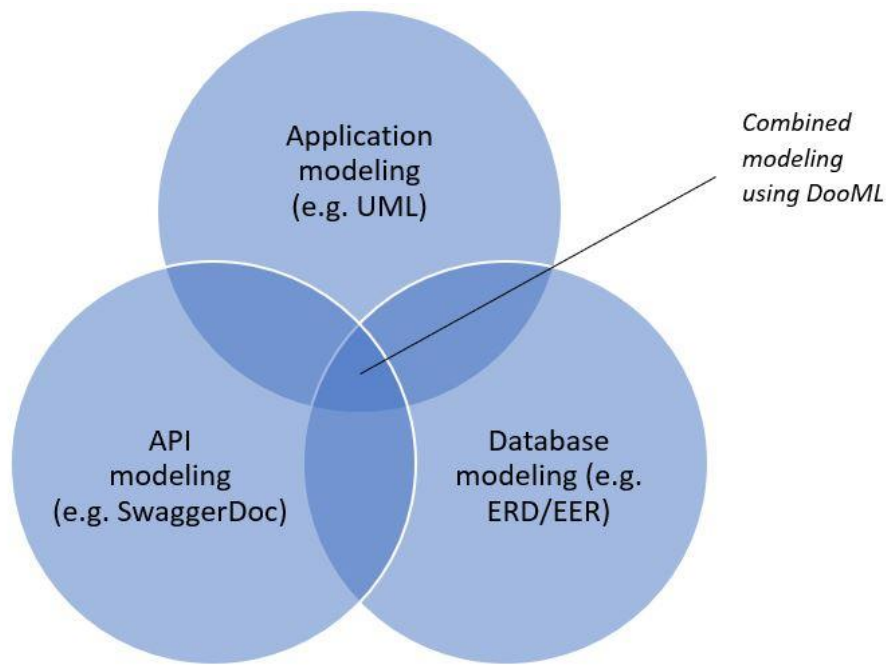

Figure 1. Venn-diagram showing the overlap between database, application and API development and corresponding modeling techniques

Using a programming language and database independent modeling scheme and syntax, a technical design for both the (code classes of) the application as well as the underlying database tabular schema can be created in one combined diagram: the DooML Archetype Diagram. By creating a combined diagram, one can use one diagram to effectively create both an application schematic as well as the corresponding database lay-out.

DooML Archetype Diagrams do not qualify as an Object Relational Mapping because it does not focus primarily on the mapping between (instantiated) objects and database relationships, however DooML Archetype Diagrams can be used as a visual schematic to ensure that objects are relationally mapped according to ORM diagram standards [9].

DooML can be construed as a Multi-Modeling language , a modeling language that consists and conforms to a combined set of metamodels, where these metamodels can be divided in sublanguages [10]. According to Boronat, Knapp, Meseguer and Wirsing, "a useful division of labor is supported by the multi-modeling language, so that some team members may concentrate their efforts on building and validating models mostly in a given sublanguage. If the team is well-coordinated and the multi-modeling language has a good infrastructure, team members working in different sublanguages will benefit from the efforts of their colleagues working in other sublanguages.". This is however not the intent of working with DooML. On the contrary, mastering a multi-modeling language as an individual, would benefit the consistency of the (software) project. We envision designers and developers working with a group of architects that simultaneously use DooML.

As the limits and boundaries are determined of what DooML is limited to and the context it resides in, the characterics of DooML and the DooML Archetype Diagram will be described.

In the upcoming chapter, the structural model (name stack, field stack and method stack) are introduced. Chapter 3 (DooML Archetype Diagram: the relational model) discusses the relational stack, relationships between xObjects/xArchetypes and the referential integrity of such constructs.

## 2. DOOML ARCHETYPE: THE STRUCTURAL MODEL

This paragraph describes the structural model and its components, responsible for demonstrating the code aspects of the modeling language. In the DooML Archetype Diagram, the structural model is responsible for displaying the properties and methods is called the property stack and the method stack. The property stack tells the developer which attributes are accessible on the object level. The method stack tells the developer which methods should be written in order for the object to function.

### 2.1. Structural Components of The DooML Archetype Diagram

The DooML Archetype Diagram is represented in a two-dimensional capacity: the first dimension describes the structural components, hence code. This structure will be addressed in this paragraph. The second dimension, representing the relational components, will be addressed in the next chapter (3).

The visual representation of the Archetype xObject as seen in figure 2 bears resemblance to a UML Class Diagram [4] but is different in a few ways. Firstly, the Field Stack shows properties or attributes that exists in code (C-) as well as on a table level (D) or on both code and table level (B). Another option is that the field (or property/attribute) could exist as an API endpoint's resource property (A). The letters can be combined to get an accurate description of the Archetype field or method. An overview of possibilities is given in figure 3.

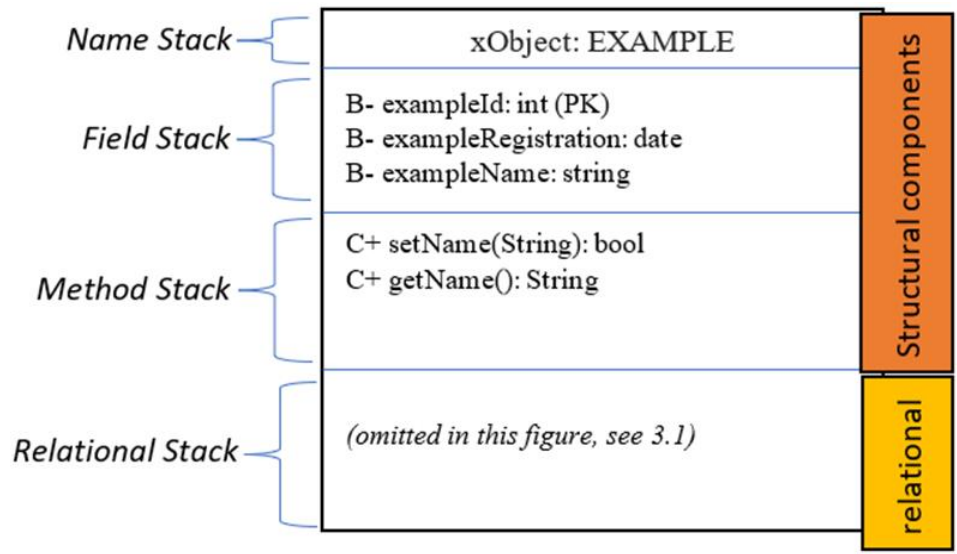

Figure 2. Example of an Archetype xObject showing the several structural components (relational stack omitted in this figure, refer to 3.1)

```
Notation of public, protected, private
+ public
-  private
# protected

Notation of variables and classes lifelines:
D - exists only in Database (table)
C - exists only in Code (class)
B - exists in Both or all possible implementations; database (table) and code (class) and/or endpoint (API)
A – exists as an API object (or: endpoint) → REST Representational State Transfer
x – exists for purpose of draft / concept versions, added for clarity
```

Figure 3. Overview of DooML Archetype or DooML Object identifiers before each property/attribute

Another difference with the UML notation is the possibility to show the same identifiers before each method or function in order to determine whether this method or function exists as a database function (D) e.g. user-defined functions in Transact-SQL, a code (C) method (for example a Java method, C# method or PHP function) or both (B). As contemporary software systems often include an API, DooML offers support for the specification of API endpoints as well by using the 'A' prefix classifier.

## 2.2. Converting from DooML to Object-Oriented Code (Conversion Rules)

Converting from a DooML Archetype Diagram to object-oriented code adheres to 7 conversion rules:

1. For each Archetype in the Archetype Diagram, a class definition will be created;
2. All fields of the field stack that have a 'A', 'B' or 'C' prefix identifier will be converted to properties of a class definition;
3. All methods of the method stack will be converted to methods/functions in a class definition;

International Journal of Software Engineering & Applications (IJSEA), Vol.13, No.5, September 2022

4. A field that has the denominator PK (Primary Key) will be considered a unique identifier (ID) or object identifier;
5. A method described with parentheses after the name declaration will be considered input (data type) for this method;
6. A method described with a data type after the colon (:), will be considered output (data type) for this method;
7. A field that has the visiblity denominator (+, -, #) transforms into a public (+), private (-) or protected (#) identifier for the corresponding class property;

## 3. DooML Archetype Diagram: the Relational Model

### 3.1. Relational Components

In 2.1 the structural segment of the DooML Archetype Diagram has been introduced. For purpose of clarity, this chapter discusses the relational segment of the Archetype. The relational segment is not limited to relationships in terms of database functionality but also covers the relationships that classes or objects (when instantiated) can be bound to.

Figure 4 shows the relational dimension of the DooML Archetype Diagram. The relational stack is an additional part of the diagram that gives the reader information on the constraints from the point of view of the database. The relational stack contains information about foreign keys, cardinality and partial or total specialization of the specific keys.

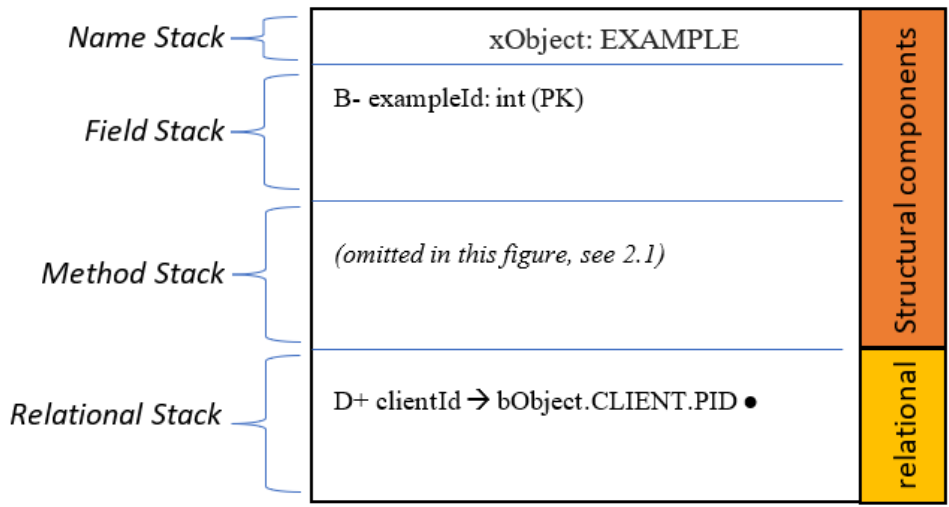

Figure 4. Example of an Archetype xObject showing the several relational components (method and field stack omitted in this figure, refer to 2.1)

Usually, the Relational stack within the DooML Archetype Diagram is used to describe which foreign key references which primary key. In figure 4 we see the sample description 'D+ exampleId → bObject.CLIENT.PID ●', which gives us information about the 'clientId' which references 'PID' in the 'bObject CLIENT'. The '●'-symbol indicates that this relationship, or more specifically the reference key connecting these two Archetypes, can not be empty and needs to be filled (hence the filled circle symbol), and therefore can not be null (NOT NULL). A open circle symbol would indicate the possibility that a relationship can be NULL. This would influence the cardinality of the corresponding relationship between xObject EXAMPLE and

bObject CLIENT in this case: the relationship would be TOTAL from the originating Archetype's point-of-view. Showing cardinality and totality in a DooML diagram does not differ much from earlier ERD implementations, as compared by Song, Evans & Park [1, p. 12]. In DooML relationship types, cardinality can be declared using the letters 1 (one-to), n (many-to in a binary relationship) or m (many-to in a ternary relationship).

Relationship types between Archetypes in DooML can be drawn with connecting lines between the Archetypes. Depending on the type of xArchetype, the line has a different drawing style.

### 3.2. Converting from DooML to a Database Schema (Conversion Rules)

Converting from a DooML Archetype Diagram to object-oriented code adheres to 6 conversion rules:

1. A many-to-many (n-to-m) relationship between 2 Archetypes will be converted to three database tables, two for the Archetypes and one for the relationship between these two Archetypes. In the (third) database table that originates from this relationship, foreign keys will be included that reference both Archetypes. All attributes/fields of the Archetype will be transformed to table columns;
2. For a one-to-many (1-to-n) or many-to-one (n-to-1) relationship, this will not be transformed to a separate database table. The Archetype connected to the 1-side will be transformed to a database table that contains all relationship fields (attributes) and an additional foreign key that references the Archetype on the n-side. The Archetype on the n-side of the relationship does not gain any additional foreign keys, and does not change in any special manner. All attributes/fields of the Archetype will be transformed to table columns;
3. For a 1-to-1 relationship, no additional table will created. One of the Archetypes is converted into a table that contains all attribute types of the relationship and a reference key to the other Archetype. This reference key in the table has a uniqueness restriction. The other Archetype remains unchanged;
4. When a table with a reference key fully participates in a relation, the reference key in that table must be filled in;
5. When a table without a reference key totally participates in a relationship, all key values must appear at least once as the value of the reference key in the other table;
6. A field that has the denominator PK (Primary Key) will be considered a unique identifier (ID) or object identifier. Depending on the database query language, this primary field can be considered UNSIGNED and bearing the AUTOINCREMENT property;

### 4. DOOML BY EXAMPLE: THE ARCHETYPE DIAGRAM

Fig. 5 shows a simple structure schema for the code and database table structure of a client-reservation system for a restaurant. The CLIENT can make a RESERVATION. Both CLIENT and RESERVATION exist as classes in code, as well as tables in the database. This combination (the existence of both database and code objects) can be modeled as an Archetype Diagram. The Archetype Diagrams consists of several Archetypes, which in turn can be instantiated as objects of a certain type. The generic version of this object in DooML is called the xObject (when instantiated) or xArchetype (when not instantiated). For demonstration purposes, these terms are interchangeable.

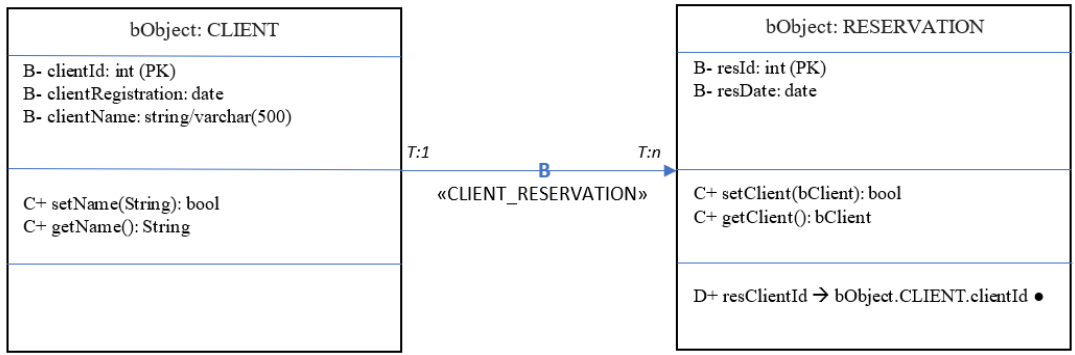

Figure 5. Simple DooML Archetype Diagram for a restaurant reservation system

When converting the bothObject (bObject) CLIENT from the DooML diagram as shown in Figure 4 to a database table (example given in SQL) and to code respectively (example given in Java), it yields the result shown in figure 5. The Java class Client is a domain model class, defining the structural components of what information should be kept on the client visiting the restaurant. The client needs an identifier (clientId) that uniquely identifies each object (class instance) but also uniquely identifies each row in the client's database table. The 'B-' classifier prefixes the names of the properties clientId, clientRegistration and clientName where the 'B' (Both) within the classifier states that clientId exists as a class property but also as a database field / column. The '-' (private) within the classifier states that on class level, this variable is not publicly accessible outside of the class or object scope.

After the name of this Archetype Field, the datatype is shown. This datatype, dependant on the letter of the prefix classifier (e.g. 'B') is the datatype that can be used in the database context (column data type) as well as in a code context (class property data type). If there is a mismatch between data types in code versus data types in database, then the code datatype precedes the database datatype, e.g. B- clientName: string/varchar(500), where string represents the datatype in code and varchar(500) represents the datatype in the database.

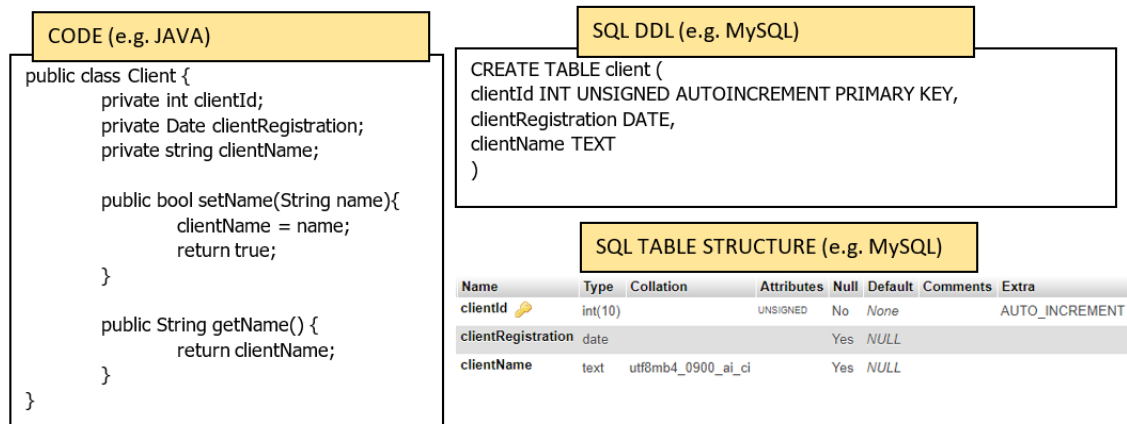

Figure 6. Example of the converted DooML Archetype Diagram to code (Java), SQL DDL (for MySQL) statement and resulting database table (in MySQL) of Object CLIENT for the restaurant reservation system

When converting the bObject RESERVATION from the DooML diagram as shown in Figure 1 to a database table (example given in SQL) and to code (example given in Java), it yields the result shown in figure 3. To ensure consistent conversion throughout several deviating use cases, a formal set of DooML conversion rules has been used for the conversion from Archetype

Diagram to the class code and SQL definition statements. The conversion rules can also be found in paragraph 2.2 and 3.2.

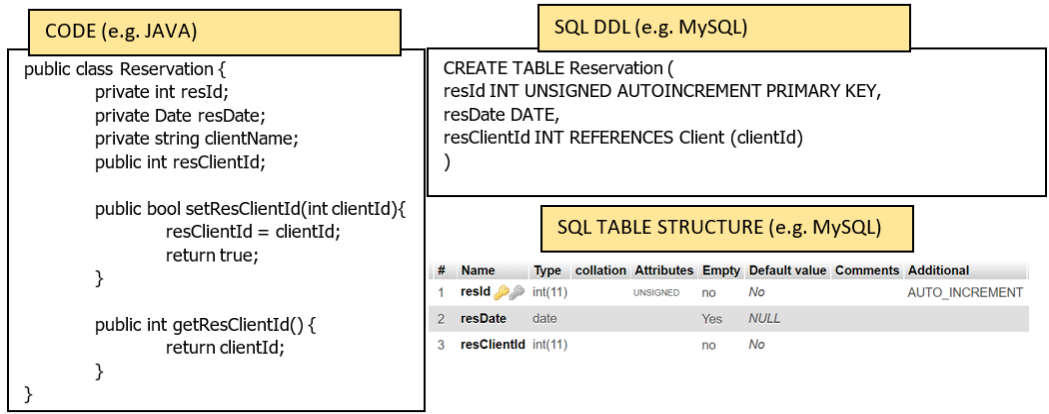

Figure 7. Example of the converted DooML Archetype Diagram to code (Java), SQL DDL (for MySQL) statement and resulting database table (in MySQL) of Object RESERVATION for the restaurant reservation system

## 5. DISCUSSION

When developing a new modeling language, questions arise, e.g. whether DooML is different or expandable to a certain other modeling language. For corresponding modeling languages one can argue that DooML is interchangeable when an XML-based specification is used to distribute the Archetype Diagrams' main characteristics. When using such an approach, it begs to differ whether a common denominator can be found to make sure that the original intended design is transferred into the respective modeling languages' constructs. Another question rises whether DooML is sustainable as a modeling language that's based on current developments in the field. For example, with the rise of NoSQL applications and non-traditional (non-relational) databases, further research is demanded on the suitability of DooML for distributed databases. Furthermore, more complex concepts of referential integrity are withheld from the DooML specification, therefore the DooML specification is receptive to improvement. Arguably, DooML Archetype Diagrams focus on the static (structural) design of an archetype. The behavior of the Archetype, analogue to the interaction between user and machine, is not yet documented. We welcome others to build upon the structural components and expand the functionality of DooML by not limiting to only classes and the importance of classes so that object lifecycles and method lifelines can be implemented. A final comment can be made on the conversion from already existing applications to DooML, whether those are databases, application programming interfaces or actual codes (so called reverse-engineering design generation).

## 6. CONCLUSIONS

DooML can be used to create stable and combined database and object-oriented software engineering diagrams. The benefit of using DooML is that one can model both the structure as well as the data storage layer of an application. The downside of using DooML is the lack of functionality to preserve referential integrity from a database perspective. Also, while DooML can be seemingly effective for large-scale software systems, for small-scale software applications the modeling phase can be a hassle to the software architect because of unnecessary overhead.

International Journal of Software Engineering & Applications (IJSEA), Vol.13, No.5, September 2022

ACKNOWLEDGEMENTS

The author would like to thank all contributors, critical friends and colleagues.

AUTHORS

**Thijs Otter** works at the Inholland University of Applied Sciences (Haarlem, The Netherlands) in the Faculty of Engineering, Design and Computing as a Lecturer in the Information Technology curriculum. He focuses on teaching (object-oriented) programming, software modeling with UML and database modeling with EER (Extended Entity Relationship) modeling. 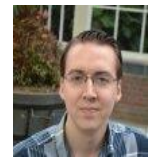